\documentclass[entropy,article,submit,moreauthors,pdftex,12pt,a4paper]{mdpi} 

\lastpage{x}
\doinum{10.3390/------}
\pubvolume{xx}
\pubyear{2013}
\history{Received: xx / Accepted: xx / Published: xx}
\pdfoutput=1

\usepackage{amssymb}
\usepackage{graphicx}

\Title{Low-temperature behaviour of social and economic networks}

\Author{Diego Garlaschelli $^{1,}$*, Sebastian E. Ahnert $^{2}$, 
Thomas M. A. Fink $^{3}$ and Guido Caldarelli $^{4,3,5}$}

\address{%
$^{1}$ Lorentz Institute of Theoretical Physics, University of Leiden, Niels Bohrweg 2, 2333 CA Leiden, The Netherlands\\
$^{2}$ Theory of Condensed Matter, Cavendish Laboratory, University of Cambridge, 
JJ Thomson Avenue, CB3 0HE Cambridge, United Kingdom\\
$^{3}$ London Institute for Mathematical Sciences, 22 South Audley St, W1K 2NY London, United Kingdom\\
$^{4}$ IMT Alti Studi Lucca, Piazza S. Ponziano 6, 55100 Lucca, Italy\\
$^{5}$ ISC-CNR, Dipartimento di Fisica, Universit\'a La Sapienza, P.le A. Moro 2, 00185 Roma, Italy}

\corres{garlaschelli@lorentz.leidenuniv.nl}

\abstract{
Real-world social and economic networks typically display a number of particular topological properties, such as a giant connected component, a broad degree distribution, the small-world property and the presence of communities of densely interconnected nodes.
Several models, including ensembles of networks also known in social science as Exponential Random Graphs, have been proposed with the aim of reproducing each of these properties in isolation.
Here we define a generalized ensemble of graphs by introducing the concept of \emph{graph temperature}, controlling the degree of topological optimization of a network.
We consider the temperature-dependent version of both existing and novel models and show that all the aforementioned topological properties can be simultaneously understood as the natural outcomes of an optimized, low-temperature topology. 
We also show that seemingly different graph models, as well as techniques used to extract information from real networks, are all found to be particular low-temperature cases of the same generalized formalism. One such technique allows us to extend our approach to real weighted networks.
Our results suggest that a low graph temperature might be an ubiquitous property of real socio-economic networks, placing conditions on the diffusion of information across these systems.
}

\keyword{Complex networks; graph ensembles; graph temperature}

\PACS{89.75.Hc, 89.75.Fb, 05.70.-a}

\begin{document}


\section{Introduction}

Complex networks have attracted the interest of physicists because statistical physics has proven to be an effective tool for the measurement and explanation of robust empirical properties of these networks \cite{barabba,siam}. 
Social and economic networks in particular often exhibit particular topological properties such as the presence of a \emph{giant connected component} (a set of mutually reachable vertices spanning a finite fraction of the system), the \emph{`small-world'} property (the combination of a large density of triangles and a short distance among nodes), \emph{community structure} (a subdivision of the network into modules of densely interconnected nodes), and a \emph{broad degree distribution} (the presence of many more highly connected vertices than expected in random graphs).

Taken together, these properties place conditions on the diffusion of information in social networks.
For instance, the \emph{`strength of weak ties'} effect \cite{granovetter}, i.e. the phenomenon by which links connecting different communities are sparser and weaker than intra-community links, implies that the dynamics of information is mostly confined within communities, with rare (but crucial in terms of network-wide communication) jumps across different communities.
In combination with the `small-world' property, this means that real social networks display what has been called the \emph{`small but slow world'} effect \cite{smallbutslow}: even if, at a purely topological level, it only takes a few steps to connect two randomly chosen nodes (that are in general found in different communities), these steps are not the ones taken by typical dynamical processes (which are instead confined within communities). 
The result is an overall slowing down of the dynamics.
 
Among the several approaches that have been explored in order to reproduce the structure of real social and economic networks, an interesting class of models has a long tradition in social network analysis and goes under the name of \emph{Exponential Random Graphs} (ERGs) \cite{holland}.
ERGs allow one to specify a set of desired topological properties or `constraints', and estimate the probability to be assigned to every possible network in order to reproduce the observed values of such properties.
More recently, ERGs have been shown to be equivalent to a class of statistical ensembles of graphs \cite{newman_expo} and have been further generalized \cite{berg,burda,mygrandcanonical,mybosefermi,ginestra,mymethod,mywreciprocity,mytriadic,mydistance}. 
This powerful formalism allows one to treat in a unified fashion a large class of models, including random graphs \cite{barabba,siam}, the Configuration Model \cite{newman_conf}, Hidden Variable models \cite{fitness,pastor} and extensions of them \cite{newman_expo,mygrandcanonical}. 
If we restrict ourselves to binary networks with a fixed number of vertices $N$ and with no self-loops or multiple edges, the analogy with statistical physics lies in the fact that in these models each link can be regarded as a `particle' that can be placed between any two vertices, subject to the constraint that the `occupation number' for each pair of vertices $i,j$ can only be $a_{ij}=0,1$ (for missing or existing links respectively) as in the familiar Fermi statistics. Clearly, $a_{ij}$ coincides with the entries of the $N\times N$ adjacency matrix $A$ characterizing the topology completely. 
Each allowed adjacency matrix $A$ corresponds to a possible configuration, and the set of possible configurations (each with its statistical weight $P_A$) defines the statistical ensemble of graphs. 
Thus, the framework of ERGs allows one to develop the `statistical mechanics of networks' by exploiting a range of tools which are well known in physics \cite{newman_expo}. 

A multitude of specifications of ERGs have been proposed in order to reproduce, mostly in isolation from each other, the topological properties mentioned above \cite{berg,burda,mygrandcanonical,mybosefermi,ginestra,mymethod,mywreciprocity,mytriadic,mydistance}.
Here we show that, if the analogy with statistical physics is completed via the introduction of the concept of \emph{graph temperature}, all the above empirical properties can be easily understood as the consequences of a single phenomenon: the fact that real networks tend to have a low value of the temperature, presumably as the result of a topological optimization driven by the cost of establishing links.

In particular we find that many well-known topological properties such as the presence of a giant component, a scale-free degree distribution, the small-world effect, and a modular or `community' structure can be easily understood in terms of the low-temperature behaviour of real networks.

\section{Temperature-dependent ensembles of graphs}

We will now generalize the existing statistical formalism in order to include the concept of graph temperature. 
To realize why this extension is important, we note that in all ERG models the probability $P_A$ depends on the \emph{energy} $E_A$, representing the topological `cost' of realising the particular graph $A$. 
The energy $E_A$ is chosen to be a linear combination of the so-called `constraints' \cite{newman_expo,mymethod}, i.e. the topological quantities that one desires to enforce. 

Now, the concept of topological cost, or energy, is unclear without the quantification of its relative 
role with respect to the available \emph{resources} that can be exploited to 
form the network. 
The relative importance of `cost' and `available resources' is usually controlled in statistical physics by the temperature.
The zero-temperature regime corresponds to complete optimization, so that only the cheapest configuration can be formed and the units of the system occupy the states with lowest energy (this is the \emph{optimized} case). 
In the opposite, infinite-temperature extreme the system does not distinguish between energetically cheap and expensive states, so that all configurations occur with the same probability.
The formalism that we develop here is particularly suitable to model networks subject to such economic/engineering constraints. We shall complement the standard results obtained in the literature for the generic finite-temperature case (which is recovered when $T=1$) with the 
interesting ones corresponding to zero and 
infinite temperature, which are not accessible to current finite-temperature models. 
We find that a range of interesting results can be obtained by even the simplest models when $T$ is allowed to vary, in particular when $T$ approaches zero. 

Our approach works equally well for directed and undirected graphs, but for the sake of simplicity we write all the expressions for the undirected case only. The generalization to directed graphs is straightforward.
Similarly, the full generalization to weighted networks is possible using the available results for weighted ensembles \cite{mybosefermi,mymethod,mywreciprocity}, but we will not consider it here. 
Rather, we will consider a partial generalization to weighted networks by exploiting one particular approach that we have recently proposed to relate edge weights to edge probabilities \cite{sebsensemble}.

\subsection{General formalism}

The most general statistical ensemble for an equilibrium undirected network is a \emph{grandcanonical} one  with $2^{N(N-1)/2}$ graphs having a fixed number of vertices $N$ and a varying number of links $L_A=\sum_{ij}a_{ij}$, controlled by the \emph{chemical potential} $\mu$ \cite{newman_expo,mygrandcanonical}.
The chemical potential is an important parameter governing the link density of the network and the probability of connections \cite{mygrandcanonical}. However, in order to have a global parameter coupled not only to the number of links, but also to any other topological property of the network, we also introduce the \emph{graph temperature} $T$.
We therefore define a generalized ensemble where the probability of graph $A$ is given by
\begin{equation}
P_A=\frac{1}{\mathcal{Z}}\exp\left[\frac{\mu L_A-E_A}{T}\right]
\label{P_general_t}
\end{equation}
where $E_A$ is the \emph{energy} of the particular graph $A$ (a function of one or more topological properties of $A$, to be specified in each particular model) and
\begin{equation}
\mathcal{Z}\equiv\sum_{A}\exp\left[\frac{\mu L_A-E_A}{T}\right]
\end{equation} 
is the \emph{grand partition function} of the ensemble.
Note that when $T\to\infty$ we have $P_A=2^{-N(N-1)/2}$ for all graphs, while when $T\to 0$ we have $P_A=1$ for the graph with the maximum value of $\mu L_{A}-E_{A}$ (or $P_A=M^{-1}$ if there are $M$ degenerate such graphs), and $P_A=0$ for all other graphs. 

The temperature in eq.(\ref{P_general_t}) might appear to be redundant, since the parameter $T$ can be in principle reabsorbed in a redefinition of $E_A$ and $\mu$ without loss of generality. 
In other words, all choices of parameters that lead to the same values of $E_A/T$ and $\mu/T$ will generate indistinguishable results, meaning that the value of $T$ is indeterminate. 
While this is mathematically true, there is a definite `physical' benefit in including the temperature as an additional parameter.
As we discuss below, the benefit is that of incorporating in $T$ all the `collective effects' arising in large networks, while leaving the local properties (such as the energy of a single link) well defined in the thermodynamical limit.
In other words, if we have empirical information about the local link energies, this will fix the scale of the problem and define the temperature unambiguously for a given network.
We will indeed show examples when, for realistic scale-free specifications of network properties, it is possible to measure the graph temperature empirically.

\subsection{Networks with finite energy per link}

In general, being a combination of topological properties, the energy $E_A$ can be an arbitrarily complicated function of the adjacency matrix $A$, but throughout the present paper we consider the simple and instructive case, explored in many models, where it can be written as a sum over the individual link energies $\epsilon_{ij}$ \cite{newman_expo,mygrandcanonical,mymethod}: 
\begin{equation}
E_A \equiv \sum_{i<j} \epsilon_{ij}a_{ij}
\label{eq:local}
\end{equation}
As we show below, this choice can - despite its simplicity - give rise not only to random graphs, but also to complex scale-free networks, small-worlds, networks with correlations, clustering, and community structure. 
The partition function reads
\begin{equation}
\mathcal{Z}=\sum_{\{A\}}\! \prod_{i<j} e^{(\mu- \epsilon_{ij})a_{ij}/T}\!=
\prod_{i<j} \left[1+e^{(\mu- \epsilon_{ij})/T}\right]
\end{equation}
and the graph probability is 
\begin{equation}
P_A=\prod_{i<j} p_{ij}^{a_{ij}}(1-p_{ij})^{1-a_{ij}}
\end{equation}
where
\begin{equation}
p_{ij}(T)=\frac {1}{e^{(\epsilon_{ij}-\mu)/T}+1}
\label{eq:p}
\end{equation}
is the probability that a link between $i$ and $j$ exists. Equation (\ref{eq:p}) has the usual form of Fermi statistics (alternative derivations of the above expression for $p_{ij}$ are given in refs.\cite{newman_expo,mygrandcanonical,newman_origin} for $T=1$).
Therefore the additivity of $E_A$ implies that each link is drawn independently with probability $p_{ij}$. 

If the form of $\epsilon_{ij}$ is further simplified, many important network models are obtained as particular cases of eq.(\ref{eq:p}), including hidden--variable models, the configuration model and random graphs \cite{newman_expo}. We shall introduce the temperature-dependent version of these models in what follows. We shall also exploit eq.(\ref{eq:p}) to study a temperature-dependent small-world model, a model with community structure, and ensembles of binary graphs derived from real-world weighted networks. Therefore eq.(\ref{eq:p}) gives rise to a rich phenomenology and will be of central importance throughout the paper.\\

Before considering particular cases, let us first note some general properties of eq.(\ref{eq:p}). Note that, independently of $T$, $p_{ij}>1/2$ when $\epsilon_{ij}<\mu$ and $p_{ij}<1/2$ when $\epsilon_{ij}>\mu$. It is interesting to consider the infinite- and zero-temperature limits, as well as the `classical' one. 

When $T\to +\infty$, eq.(\ref{eq:p}) implies that 
\begin{equation}
p_{ij}(+\infty)=\frac{1}{2}
\label{ti}
\end{equation}
irrespective of the values of $\epsilon_{ij}$ and hence of the differences in the 
cost of links. As a consequence, the network is a random graph with $p=1/2$ and is therefore trivial. Note that in this case any two configurations $A$ and $B$ become equiprobable ($P_A=P_B$).

When $T=0$, we instead have
\begin{equation}
p_{ij}(0)=\Theta(\mu-\epsilon_{ij})
\label{eq:theta}
\end{equation}
where $\Theta(x)=1$ if $x>0$ and $\Theta(x)=0$  if $x<0$. Technically, we should define $\Theta(0)=1/2$ in order to capture the correct behaviour of eq.(\ref{eq:p}), even if we will not encounter this situation in what follows.
The above equation means that only those pairs of vertices for which $\epsilon_{ij}<\mu$ are connected. 
This is analogous to the well-known \emph{degenerate} behaviour of Fermions at zero temperature, and $\mu$ is also termed the \emph{Fermi energy} $\epsilon_F=\mu$. 
This clarifies the role of $\mu$ as the available energy per link when $T\to 0$: at absolute zero, only the topology with the minimum value of $E_A -\mu L_A$ can be realized. This topology is obtained by drawing all and only the links with $\epsilon_{ij}<\mu$.

A final general comment is that eq.(\ref{eq:p}) reduces to the `classical limit' \cite{newman_expo}
\begin{equation}
p_{ij}(T)\approx e^{(\mu-\epsilon_{ij})/T}
\qquad \textrm{when}\quad e^{(\epsilon_{ij}-\mu)/T}\gg 1
\label{classical}
\end{equation}
We will consider the above limit in some applications later on.\\

An important consequence of the above general considerations is that, since both $\epsilon_{ij}$ and $\mu$ are link-specific quantities, we will be interested in the case when their value is finite and independent of the network's size $N$. 
This is because large-scale networks arise as a bottom-up combination of local link formation processes. 
The energy per pair of vertices should therefore have a finite value in the thermodynamic limit ($N\to\infty$). 
For the same reason, we will assume that $\epsilon_{ij}$ and $\mu$ are independent of the temperature $T$. 
In other words, $\epsilon_{ij}$ and $\mu$ fix the scale of pair-specific properties (that should remain well-defined in the large $N$ limit), and the collective (network-wide) effects are reabsorbed in $T$.
As we have anticipated above, this is the main added value of isolating $T$ from the other parameters of the model, and the ultimate reason why we believe that investigating the temperature dependence of network ensembles is important.
As a final remark, we require $\epsilon_{ij}$, $\mu$ and $T$ to be dimensionless. If we imagine that $\epsilon_{ij}$ is (a function of) an empirically measurable quantity such as distance or money, an dimensionless specification can be achieved by assuming that both $\epsilon_{ij}$ and $\mu$ have been preliminarily divided by some appropriately averaged (either over vertices or vertex pairs) value of $\epsilon_{ij}$, and by simply considering $T$ as a dimensionless parameter. We will discuss this point in each of the following examples.

\section{Random Graphs: vanishing of the percolation threshold at zero temperature\label{sec:rg}}

In what follows we consider various specific cases. 
The simplest scenario is when all link energies are equal: $\epsilon_{ij}=\epsilon$. 
This yields a temperature-dependent random graph of the Erd\"os-R\'enyi type, since all probabilities $p_{ij}$ are equal to 
\begin{equation}
p(T)=\frac{1}{1+e^{(\epsilon-\mu)/T}}
\end{equation}
Note that, if we assume that the $\epsilon$ has been divided by its average value over all pairs of vertices to make it dimensionless, we should simply set $\epsilon=1$. 
When looking at the above formula, as well as the following ones, this is the value of $\epsilon$ that we should keep in mind. 

While the properties of the random graph are well known, in our framework some intriguing results emerge as the temperature is varied, and in particular when $T\to 0$. 
First of all, we note that 
\begin{equation}
p(0)=\Theta(\mu-\epsilon)
\end{equation}
implying that the graph is either fully connected ($\mu>\epsilon$) or empty ($\mu<\epsilon$).\footnote{Technically, we recall that if $\mu=\epsilon$ then $p(0)=1/2$, i.e. the graph is half-connected.}
This result provides us with a useful (for our purposes in what follows) definition of `sparseness' of a network. We define a random graph as \emph{sparse} (\emph{dense}) if $\epsilon>\mu$ ($\epsilon<\mu$), since when $T\to 0$ the graph becomes empty (fully connected). 
This means that, at finite temperature, a sparse graph (as defined above) will be such that $p(T)<1/2$ and a dense graph will be such that $p(T)>1/2$. At infinite temperature, both sparse and dense graphs converge to the intermediate density $p(+\infty)=1/2$.

\subsection{Critical percolation temperature}

Before considering other models, it is quite interesting to consider the percolation transition marking the onset of a giant connected component in an infinitely large random graph.
For random graphs, it is well known that this transition occurs when the connection probability $p$ is set to the critical value $p_c\sim 1/N$, i.e. when the function $f(N)$ introduced above is $f(N)\sim N$.
In our framework, since $\epsilon$ and $\mu$ are fixed, we can regard the phase transition as temperature-dependent.
If $\epsilon>\mu$ then $p(T)>1/2>p_c$ at all temperatures, meaning that dense graphs are obviously always above the critical threshold.
If $\epsilon<\mu$, there is a \emph{critical percolation temperature} $T_c$ such that $p(T_c)=p_c\sim 1/N$. Inverting, we find that for sparse graphs
\begin{equation}
T_c(N)\sim\frac{\epsilon-\mu}{\ln N}\qquad \epsilon<\mu
\label{tcfinite}
\end{equation} 
In the thermodynamic limit, we have
\begin{equation}
\lim_{N\to\infty}T_c(N)= 0
\label{eq:Tc}
\end{equation} 
meaning that when $N\to\infty$,  
\emph{the critical percolation temperature tends to zero, i.e. the zero-temperature topology naturally sets at the critical point $p=p_c$}. 
It is well know that at this critical point the distribution of the sizes of connected components of the network has a power law distribution of the form $P(s)\propto s^{-5/2}$ \cite{barabba}.
Interestingly, this behaviour is similar to a scenario explored in the theory of Self-Organized Criticality (SOC), where the onset of the SOC behaviour has been related to the vanishing of the critical temperature \cite{soc}. 
Combining together the above results about dense and sparse graphs, we find that, \emph{irrespective of their density, at finite temperature infinitely large random graphs are always above the percolation threshold}.

\subsection{Large and sparse graphs have low temperature}

We note that the link density of most real-world socio-economic networks is (significantly) smaller than $1/2$. This means that, when modeled as random graphs (i.e. when considering a connection probability $p$ equal to the observed link density $f$), real networks systematically fall in the `sparse graph' category and are therefore such that $\epsilon>\mu$.
It should also be noted that in most cases the observed density typically decays as $1/f(N)$ where $f(N)$ is an increasing function of $N$. 
This means that, in order to reproduce the empirically observed density, random graphs should be such that 
\begin{equation}
\frac{1}{1+e^{(\epsilon-\mu)/T}}=\frac{1}{f(N)}
\end{equation}
which implies
\begin{equation}
T=\frac{\epsilon-\mu}{\ln[f(N)-1]}
\label{eq:low}
\end{equation}
This result shows that larger graphs have a smaller temperature, providing a first indication of the fact that large real-world networks might be generally characterized by a small value of the graph temperature.

It is also important to note that, for most observed networks, $f(N)\simeq c N$ with $c\gtrsim 1$. In combination with eq.(\ref{tcfinite}) this means that, when modeled as random graphs, large real-world networks have a low but non-zero temperature, i.e. they are `just above' the percolation threshold. This is enough to ensure that large networks have a giant connected component.
For social networks, being above the percolation threshold ensures that, starting from any node in the giant connected component, information can diffuse to any other node of the same component.
Since in the thermodynamic limit the giant component spans a finite fraction of an infinite network, this means that information can diffuse to a macroscopic scale.\\

Obviously, real networks are very different from random graphs. Still, the above considerations hold true also for more realistic models of networks displaying `scale-free' and `small-world' properties, as we show in the next examples.
On the other hand, we also know that the way  information diffuses on real social networks is not uniquely determined by whether a giant connected component exists. 
In particular, we know that information is mostly confined within denser modules, a feature that requires the network to be partitioned into so-called `communities' \cite{smallbutslow}.
We will discuss this point later on, when we will introduce a model with temperature-induced community structure.

\section{Fitness models: random graphs at high temperature, scale-free networks at low temperature\label{sec:fitness}}

Another case of great interest is when the link energy in eq.(\ref{eq:local}) is the sum of two single-vertex contributions: 
\begin{equation}
\epsilon_{ij}=\epsilon_i+\epsilon_j
\label{eq:sum}
\end{equation}
For future convenience, we assume that $\epsilon_i\le 0$ $\forall i$ (this can always be achieved by an irrelevant overall shift in the energies $\epsilon_i\to\epsilon_i- \epsilon_{max}\le 0$).
Moreover, to have a dimensionless quantity we imagine that $\epsilon_{i}$ (and similarly $\epsilon_{j}$) has been preliminary divided by the absolute value of its average over all vertices. After these operations, we therefore have $\overline{\epsilon}=-1$ where the bar denotes an average of $\epsilon_{i}$ over vertices.

The above choice leads to the graph energy
\begin{equation}
E_A=\sum_{i}\epsilon_i k_i
\label{eq:cm}
\end{equation}
where $k_i\equiv\sum_j a_{ij}$ is the degree (number of links) of vertex $i$. 
Note that all graphs $A$ with the same degrees have the same energy $E_A$ and are therefore equiprobable. 
This case therefore represents the grand-canonical version of the so-called \emph{Configuration Model}, i.e. a model of random networks with given degrees \cite{newman_expo}. 
It can also be regarded as a particular case of the class of Fitness Models \cite{fitness} where each node $i$ is characterized by a \emph{`fitness'} or \emph{`hidden variable'} $x_i$ determining the connection probability.
The novelty of our approach is that the node fitness $x_i\equiv e^{-\epsilon_i/T}$ and the `fugacity' $z\equiv e^{\mu/T}$ (in terms of which the model is conveniently described \cite{newman_origin,mywtw}) now depend on $T$. We can therefore write
\begin{equation}
p_{ij}(T)=\frac {1}{e^{(\epsilon_i+\epsilon_j-\mu)/T}+1}
=\frac{zx_ix_j}{1+zx_ix_j}
\label{eq:conf}
\end{equation}
which reduces to the random graph case discussed in the previous section when all vertices have the same value of $\epsilon_i$ or equivalently $x_i$.\\

We now consider a standard procedure to obtain scale-free degree distributions, i.e. by assigning each vertex $i$ a fitness $x_i$ drawn from a power-law distribution $\rho(x)\sim x^{-\gamma}$. It has been shown that this choice leads to a scale-free degree distribution with the same exponent $-\gamma$, followed by a cut-off for large degrees \cite{newman_origin}. The cut-off arises from the fact thar $p_{ij}$ saturates to $1$ as $x_i\to +\infty$, which is in turn a reflection of the fact that the degrees cannot exceed the maximum number $N-1$. Clear empirical evidences of this saturation have been observed for instance in the analysis of the Internet \cite{newman_origin} and of the World Trade Web (WTW) \cite{mywtw}.

To highlight the role of $T$, we now rephrase the above results in terms of the energies $\epsilon_i$. For convenience we introduce the non-negative quantity $\phi_i\equiv-\epsilon_i\ge 0$, which measures the tendency of vertex $i$ to form connections \cite{fitness}. Similarly, we define $\phi_0\equiv -\mu$. Now, if we want $x$ to be distributed according to 
\begin{equation}
\rho(x)=(\gamma-1)x^{-\gamma}
\end{equation}
(where $1\le x<+\infty$ and $\gamma>1$), then the quantity $\phi_i=-\epsilon_i=T\ln x_i$ must be distributed according to 
\begin{equation}
q(\phi)=\frac{\gamma-1}{T}e^{-\phi(\gamma-1)/T}
\end{equation}
Now, since $\phi$ does not depend on $T$, $q(\phi)$ must be $T$-independent as well. The only possibility is therefore $(\gamma-1)/T=\lambda$ where $\lambda$ is a constant independent of $T$. 
Note that the mean of the distribution $q(\phi)$ is $\overline{\phi}=\lambda^{-1}$. On the other hand, since $\phi_i=-\epsilon_i$ and $\overline{\epsilon}=-1$, we also have $\overline{\phi}=-\overline{\epsilon}=1$.
This means that we must set $\lambda=1$. This yields $\gamma=1+T$ and
\begin{eqnarray}
q(\phi)=e^{-\phi}\quad (\phi\ge 0)\label{eq_phidistr}\\
\rho(x)=Tx^{-1-T}\quad(x\ge 1)
\label{eq_Tdistr}
\end{eqnarray}
which is an important result showing how $T$ determines $\rho(x)$ and consequently the topology of the network. 

For instance, in the classical limit (\ref{classical}) we recover the $T$-dependent version of a model studied in ref.\cite{fitness}: since $p_{ij}(T)\approx zx_ix_j$, the expected degree  $\bar{k}_i=\sum_j p_{ij}(T)\approx zx_i\sum_j x_j$ is proportional to $x_i$ and is therefore distributed as 
\begin{equation}
P(\bar{k})\propto \bar{k}^{-1-T}
\label{eq:fitness1}
\end{equation}
In this case there are no degree correlations due to the factorization of $p_{ij}(T)$. 

In the more general case (i.e. outside the `classical' regime), $P(k)$ has a power-law region with an exponent that is still an increasing function of $T$, followed by a cut-off arising from the saturation of $p_{ij}(T)$. The power-law region narrows as $T$ increases.
This qualitative behaviour can be characterized rigorously by computing $\bar{k}_i$ as a function of $x_i$ or $\phi_i$, and inverting this relation to find $P(\bar{k})$ from $\rho(x)$ or $q(\phi)$.
This is not easy in general, but here we show that in the three paradigmatic cases $T=+\infty$, $T=1$ and $T=0$ it can be done successfully. 

\subsection{High-temperature regime ($T=+\infty$)}

For $T=+\infty$ we have the usual result $p_{ij}(+\infty)=1/2$. So the network is a random graph with density $1/2$ and average degree $N/2$. In this regime the degree distribution $P(\bar{k})$ approaches a trivial Poisson distribution with mean $N/2$. There are no degree-degree correlations, and all nodes have an expected clustering coefficient equal to $1/2$.

\subsection{Finite-temperature regime ($T=1$)}

For $T=1$, denoting $p_{ij}(T)=p(\phi_i,\phi_j)$ the expected degree of a vertex with fitness $\phi$ can be evaluated as the integral 
\begin{eqnarray}
\bar{k}&=&
N\!\int_0^{+\infty}\! p(\phi,\phi')q(\phi')d\phi'
=N\!\int_0^{+\infty}\! \frac{q(\phi')}{e^{\phi_0-\phi-\phi'}+1}d\phi'\nonumber\\
&=&N \frac{\ln(e^{\phi_0-\phi}+1)}{e^{\phi_0-\phi}}
=Nzx\ln\frac{1+zx}{zx}
\label{t1}
\end{eqnarray}
which is an increasing function of $x$ and is therefore invertible.
If $x(\bar{k})$ denotes the inverse function, the expected degree distribution is $P(\bar{k})=\rho[x(\bar{k})]dx/d\bar{k}$. 
Note that $\bar{k}\propto x$ for small $x$, while $\bar{k}\to N$ for large  $x$. Thus in the linear regime (small $\bar{k}$) we have $x\propto \bar{k}$ as in the classical limit, so that $dx/d\bar{k}$ is constant and
\begin{equation}
P(\bar{k})\propto \rho[x(\bar{k})]\propto \bar{k}^{-2}
\end{equation}
This scale-free region is followed by a cut-off for large $k$, corresponding to the `saturated' behaviour. 

\subsection{Low-temperature regime ($T=0$)}

Finally, when $T=0$ the expression for $\rho(x)$ in eq.(\ref{eq_Tdistr}) breaks down since all the $x_i$'s become infinite, and from eq.(\ref{eq:theta}) we find 
\begin{equation}
p_{ij}(0)=\Theta(\phi_i+\phi_j-\phi_0)
\end{equation}
Surprisingly, this coincides with another model introduced in ref.\cite{fitness}, which precisely assumes $q(\phi)=e^{-\phi}$ as in eq.(\ref{eq_phidistr}) and 
thus turns out to be the zero-temperature limit of our general model. This model is intriguing since a derivation similar to that in eq.(\ref{t1}) shows that it yields a purely scale-free degree distribution 
\begin{equation}
P(\bar{k})=(N e^{-\phi_0})\: \bar{k}^{-2}=(N e^{\mu})\: \bar{k}^{-2}
\label{eq:fitness2}
\end{equation}
(now without cut-off), even if no power-laws are introduced `by hand' in the model \cite{fitness,pastor}. 
Moreover, the model displays anticorrelation between degrees: the average nearest neighbour degree scales as 
\begin{equation}
\bar{k}^{nn}(\bar{k})\propto \bar{k}^{-1}
\label{eq:fitness3}
\end{equation}
and the clustering coefficient scales as \begin{equation}
\bar{c}(\bar{k})\propto \bar{k}^{-2}
\label{eq:fitness4}
\end{equation}
 (times logarithmic corrections) \cite{fitness,pastor}.

We note that, while in ref. \cite{fitness} the above model was proposed as an alternative way to produce scale-free networks, different from the specification leading to eq.(\ref{eq:fitness1}), here we find that both choices are actually two particular cases of the same temperature-dependent model. 
We also note that eq.(\ref{eq:fitness2}) cannot be retrieved as the zero-temperature limit of eq.(\ref{eq:fitness1}), since in such limit the `classical' approximation (\ref{classical}) is no longer valid.
Rather, the above results show that as $T$ goes to zero the exponent of the degree distribution approaches $-2$, with a gradual disappearance of the upper cut-off. 
Moreover, we stress that while the topological properties of the network depend on both the temperature $T$ and the chemical potential $\mu$, the latter strongly determines the mean of the degree distribution (i.e. the link density) but not its functional form, which is instead mainly determined by $T$.\\

Taken together, the above results lead to the following intriguing conclusion: in this model, \emph{correlated scale-free networks with exponent $-2$ naturally arise as the optimized topology at zero temperature. As $T$ grows, the correlations become weaker, the exponent of $P(\bar{k})$ increases and a cut-off appears destroying the purely scale-free behaviour, until for $T\to\infty$ the network becomes an uncorrelated random graph with a Poisson degree distribution}. 
In our framework, it is clear that $\phi_0$ plays the role of a Fermi energy. We can also interpret the correlations at $T=0$ as the collective need to minimise the total energy, an effect that gradually weakens as $T$ increases.

\subsection{The temperature of real binary networks}

We now make some important considerations about the temperature of real-world binary networks.
The degree distribution of most real scale-free networks has a broad tail of the form
\begin{equation}
P(k)\propto k^{-\gamma}
\qquad 2\lesssim\gamma\lesssim 3
\label{eq:pk}
\end{equation}
The above observed range of the exponent is another remarkable indication that real networks are consistent with a low-temperature model. 
In particular, all binary scale-free networks in the `classical regime' described by eq.(\ref{eq:fitness1}) are consistent with a temperature 
\begin{equation}
T_{binary}=\gamma-1\quad\implies\quad 1\lesssim T_{binary}\lesssim 2
\label{eq:Tbinary}
\end{equation}
Scale-free networks outside the classical regime are instead characterized by an even lower temperature, since we have shown that $\gamma=2$ is realized at $T=0$. For these networks, a small positive value $T_{binary}>0$ is already enough to produce a realistic degree distribution with $\gamma>2$.

We finally note that, if one has access to the empirical distribution $\rho(x)$, one can measure $T_{binary}$ for any real network which is well described by eq.(\ref{eq:conf}), even if this network is not scale-free. This is for instance possible for the WTW, where $x_i$ has been identified with the Gross Domestic Product of country $i$, whose distribution has a fat tail consistent with a power-law with exponent $-2$ \cite{mywtw}. This means that $T_{binary}^{WTW}\approx 1$ and that eq.(\ref{t1}) applies. This is consistent with the observed saturated behaviour of $k(x)$ and the cut-off displayed by $P(k)$ for the real WTW \cite{mywtw}.

\section{More general models}

One can further explore eq.(\ref{eq:p}) by considering different forms of $q(\phi)$ and of $\epsilon_{ij}$ as a function of $\phi_i$ and $\phi_j$, thus recovering the whole class of fitness models \cite{fitness} with generic $p_{ij}=p(\phi_i,\phi_j)$. 
An even more general case is when $\epsilon_{ij}$ cannot be written as a function of single-vertex contributions, so that each pair of vertices has an associated quantity $\phi_{ij}\equiv -\epsilon_{ij}$ drawn from a distribution $q(\phi)$, and a probability $p_{ij}=p(\phi_{ij})$ to exist. 
This corresponds to the general case defined by eqs. (\ref{eq:local}) and (\ref{eq:p}).

The vanishing of the percolation threshold as shown previously in eq.(\ref{eq:Tc}) for the random graph example is actually a more general result, and holds even when different pairs of vertices have different values of $\epsilon_{ij}$ as in eq.(\ref{eq:p}), i.e. when $\epsilon_{min}\le\epsilon_{ij}\le\epsilon_{max}<\mu$. In this case, we must have 
\begin{equation}
T_{min}(N)\le T_c(N)\le T_{max}(N)
\end{equation}
where \begin{equation}
T_{min}(N)\equiv\frac{\epsilon_{min}-\mu}{\ln N} , \qquad T_{max}(N)\equiv\frac{\epsilon_{max}-\mu}{\ln N} 
\end{equation}
From the above equations, it follows that  
\begin{equation}
\lim_{N\to\infty}T_{min}(N)= 0,\quad\lim_{N\to\infty}T_{max}(N)= 0\quad\implies\quad\lim_{N\to\infty}T_{c}(N)= 0
\label{eq:siempre}
\end{equation}
Therefore the critical temperature vanishes in this case as well.
Again, this suggests why most large social networks display a giant connected component, including the class of scale-free networks consistent with the model described above.\\

In what follows we consider three particular cases of the above model: a temperature-dependent modification of the \emph{`small-world'} network model \cite{smallworld}, a novel model of networks with community structure, and a generalization of ensembles of binary graphs derived from weighted networks \cite{sebsensemble}. The latter allows for a direct and simple extension to real weighted networks.

\section{A temperature-driven small-world model\label{sec:sw}}

The energy in eq.(\ref{eq:p}) is particularly suitable as a model of networks with geometric constraints \cite{mydistance,spatialnetworks,myfilling}. Indeed, if vertices represent points of a metric space, the distance between pairs of vertices will affect the cost of a link between them. 
We now show that this simple fact leads to a straightforward definition of a temperature-dependent model of small-world networks, that can be also extended to generate networks that are both small-world and scale-free.

\subsection{Non-scale-free small-worlds}

In the simplest situation, the energy $\epsilon_{ij}$ of a link is simply proportional to the distance $d_{ij}$ between its end-point vertices.
If we imagine that both $\epsilon_{ij}$ and $d_{ij}$ have been made dimensionless by dividing each of them by the respective average value over all pairs of vertices, the proportionality constant drops out and we can simply write 
\begin{equation}
\epsilon_{ij}=d_{ij}
\end{equation}
This implies 
\begin{equation}
E_A=\sum_{i<j}\epsilon_{ij}a_{ij}=\sum_{i<j}d_{ij}a_{ij}
\label{eq_distance}
\end{equation}
so that the probability of a link being there between $i$ and $j$ reads
\begin{equation}
p_{ij}(T)=\frac {1}{e^{(d_{ij}-\mu)/T}+1}
\label{eq_pd}
\end{equation}

\begin{figure}
\begin{center}
\includegraphics[width=.75\textwidth]{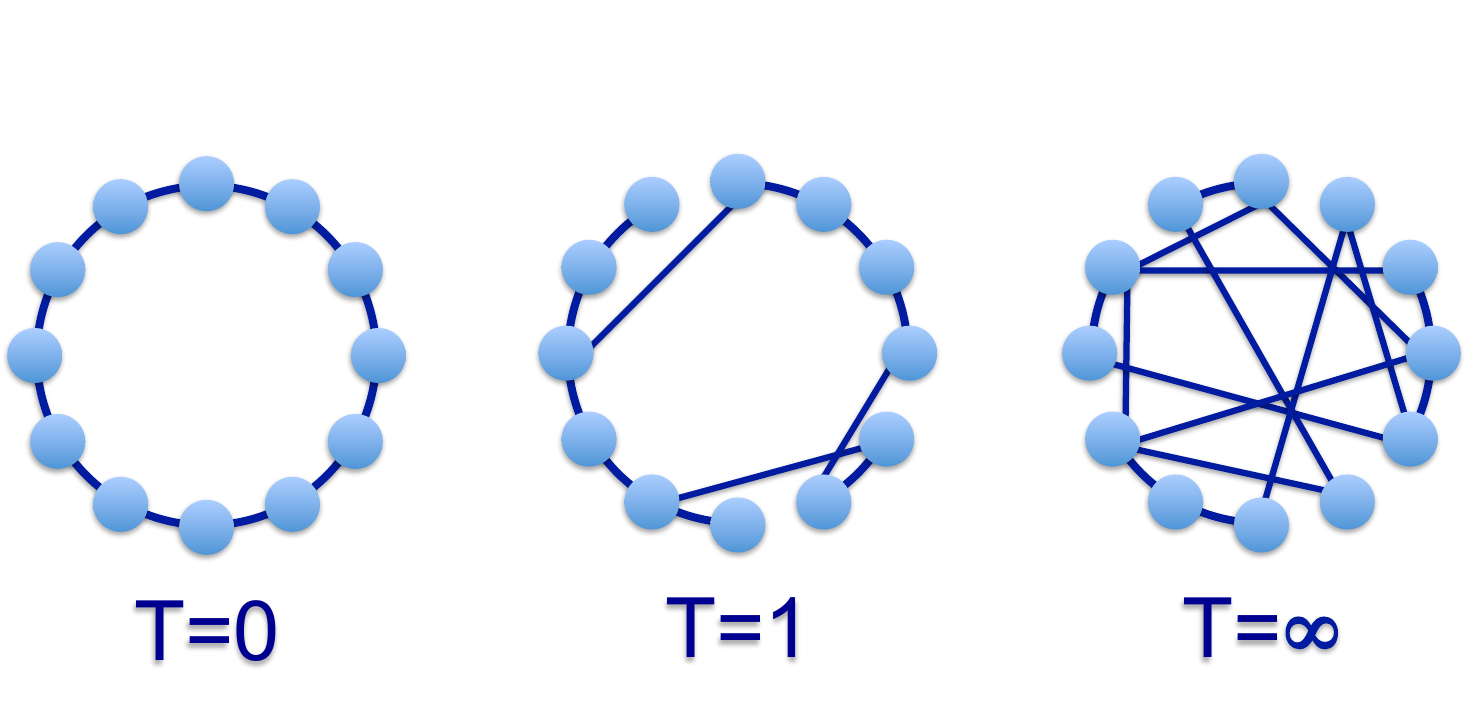}%
\end{center}
\caption{A temperature-dependent small-world model with vertices arranged in a circle and chemical potential $d<\mu<2d$ (where $d$ is the dimensionless distance between nearest neighbours along the circle). When $T=0$ (left), the network is a ring with first-neighbour interactions. When $T=\infty$ (right), the network is a random graph with connection probability $p=1/2$. When $T=1$ (center), the network is a `small-world' with a few long-range connections and an incomplete circular `backbone'.}
\label{fig:sw}
\end{figure}

Let us first consider the zero-temperature behaviour. The above probability becomes
\begin{equation}
p_{ij}(0)=\Theta(\mu-d_{ij})
\label{eq:theta2}
\end{equation}
which is nothing but the definition of a local `metric' network connecting the geometrically closest vertices (with a connectivity range set by the chemical potential $\mu$).
Consider for instance $N$ vertices equally spaced on a circle.
Let $d$ be the dimensionless distance between nearest neighbours along the circle, i.e. $d_{ij}=d$ if $i$ and $j$ are first neighbours, $d_{ij}=2d$ if they are second neighbours, and so on.
At zero temperature, eq.(\ref{eq:theta2}) implies that if $d<\mu<2d$ then the network is a ring with first-neighbour interactions (as in Fig. \ref{fig:sw}), if $2d<\mu<3d$ then the network is a ring with second-neighbour interactions, and in general if $md<\mu<(m+1)d$ (where $m$ is a positive integer) the network is a ring with $m$th-neighbour interactions.
If vertices are instead the nodes of a $D$-dimensional lattice and $d$ is the lattice spacing, then when $md<\mu<(m+1)d$ the zero-temperature network is a lattice with the same dimensionality and with $m$th-neighbour interactions.\footnote{Note that, if we allow the chemical potential to take precisely the integer value $\mu=md$, then the pairs of vertices separated by a distance $d_{ij}=md$ will be connected with probability $p_{ij}(0)=1/2$, adding a sort of `random anomaly' to the ring-like or lattice structure. For this reason, we have deliberately restricted $\mu$ to take the non-integer values $md<\mu<(m+1)d$ so that $\mu\ne md$.}

If the temperature is slightly increased from $T=0$ to a small positive value, then these regular ring-like or lattice structures will be perturbed, with a small number of short-range connections being replaced by longer-range ones (see fig. \ref{fig:sw}). 
At higher temperature, the zero-temperature structure becomes increasingly obscured and even longer-range connections are formed, until at infinite temperature the network becomes a completely random graph with connection probability $p_{ij}(+\infty)=1/2$.\\

From the above discussion, it is clear that this model is very similar to the popular `small-world' model by Watts and Strogatz (WS) \cite{smallworld}, where an initial lattice is perturbed by redirecting its links, with probability $p$, to randomly chosen vertices.
In that model, if $p=0$ the original lattice is preserved, if $0<p<1$ part of the original lattice coexists with long-range connections or `shortcuts', while if $p=1$ all the links are rewired as in a random graph.

There are two main differences between our model and the WS one. 
First, here for all finite values of $T$ longer links have a smaller probability than shorter links, whereas in the WS model distance does not affect the probability of creating shortcuts (which is a less realistic situation).
Second, here the totally random ($T=+\infty$) configuration has density $1/2$ irrespective of the density of the initial ($T=0$) lattice, whereas in the WS model the totally random ($p=1$) configuration has the same density as the initial ($p=0$) lattice.
In fact, here the density depends monotonically on $T$, while in the WS model it is independent of $p$.
Therefore the low-temperature regime of our model is similar to the WS model in the regime of low rewiring probability, whereas the behaviour of the two models differs in their high-temperature/high-rewiring regimes.

In the limit of low rewiring probability, the standard WS model exhibits the so-called `small-world' effect, i.e. the combination of a large value of the clustering coefficient (measuring the average fraction of realized triangles at each node) and of a small value of the average vertex-vertex distance (which increases only logarithmically with the size of the graph) \cite{barabba}.
Since these properties are found in the low-rewiring regime, we can expect that they would be generated also in the low-temperature regime of our model.
Again, this means that the empirically observed properties (in this case the small-world effect) are reproduced for small positive values of the graph temperature.

\subsection{Scale-free small-worlds}

As for the random graph model, we know that the simple small-world model (either the original WS one or our temperature-dependent reformulation above) does not reproduce the broad degree distribution so widely observed in real networks.
Here, we briefly discuss how our model above can be extended in order to account for a heterogeneous - and if necessary, scale-free - degree distribution.

To this end, we combine two models considered so far, by assuming that the link energy in eq.(\ref{eq:local}) is determined not only by distances, as in the above model, but also by vertex-specific properties, as in eq.(\ref{eq:sum}). This leads to
\begin{equation}
\epsilon_{ij}=d_{ij}+\epsilon_{i}+\epsilon_{j}
\label{eq:d+e}
\end{equation}
where now we imagine that $d_{ij}$ has been divided by its average over all pairs of vertices, while $\epsilon_{i}$ and $\epsilon_{j}$ have been divided by their average over all vertices.
Correspondingly, eq.(\ref{eq:local}) becomes 
\begin{equation}
E_A=\sum_{i<j}d_{ij}a_{ij}+\sum_i\epsilon_{i}k_i
\label{eq_DCM}
\end{equation}
and eq.(\ref{eq:p}) becomes
\begin{equation}
p_{ij}(T)=\frac {1}{e^{(d_{ij}+\epsilon_i+\epsilon_j-\mu)/T}+1}
=\frac{zx_ix_j e^{-d_{ij}}}{1+zx_ix_je^{-d_{ij}}}
\label{eq:pDCM}
\end{equation}
(where we have used the same definitions as in sec. \ref{sec:fitness}).
The above model has been recently exploited in order to study the spatial properties of the World Trade Web, where $d_{ij}$ is the (dimensionless) geographic distance between countries $i$ and $j$ \cite{mydistance}.\\

Clearly, a sufficiently heterogeneous distribution of the values of $\epsilon_i$ will induce a broad degree distribution, exactly as we showed in sec. \ref{sec:fitness}.
In particular, a suitable choice allows to  reproduce the scale-free and small-world properties simultaneously.
However, a general conclusion one can learn from this model is that, if the distances arise from a homogeneous spatial distribution of vertices and if the degree distribution induced by eq.(\ref{eq_DCM}) is very broad, this typically means that, while the distribution of the sums $\epsilon_i+\epsilon_j$ is very broad, the distribution of the distances $d_{ij}$ is much more narrowly concentrated around its average value $\overline{d}=1$.
Looking at eq.(\ref{eq:d+e}), this means that the distribution of $\epsilon_{ij}$ is mainly determined by that of $\epsilon_i+\epsilon_j$, i.e. we can make the approximation
\begin{equation}
\epsilon_{ij}\approx \bar{d}+\epsilon_{i}+\epsilon_{j}=1+\epsilon_{i}+\epsilon_{j}
\label{eq:approx}
\end{equation}
Clearly, the constant unit term $\bar{d}=1$ can be reabsorbed in the chemical potential by defining $\mu'\equiv \mu-1$, thus leading us back to the model defined by eq.(\ref{eq:sum}) in sec. \ref{sec:fitness}.
For instance, when $T=0$ the degree distribution reads
\begin{equation}
P(\bar{k})\approx(N e^{\mu-1})\: \bar{k}^{-2}
\label{eq:fitnesssw}
\end{equation}
and the degree correlations and clustering properties are still given by eqs.(\ref{eq:fitness3}) and (\ref{eq:fitness4}).\\

The above considerations mean that the scale-free property automatically implies the small-world one, whereas the converse does not hold in general.
An empirical confirmation of this idea comes from the recent application of eq.(\ref{eq:pDCM}) to the World Trade Web \cite{mydistance}, which showed that this model only slightly improves the fit to real data with respect to the simpler model defined by eq.(\ref{eq:conf}), and is in any case outperformed by other models which, instead of the distances, add to eq.(\ref{eq:conf}) a different piece of information.
Since for a realistically broad distribution of $\epsilon_i$ the model reduces to the distance-independent one introduced in sec. \ref{sec:fitness}, we can again conclude that the relevant range of temperature is when $T$ is small but non-zero, exactly as in the distance-independent model.
Moreover, for scale-free networks with a degree distribution given by eq.(\ref{eq:pk}) the empirical value of the temperature is still given by eq.(\ref{eq:Tbinary}).

\section{A model of networks with low-temperature community structure}

We now come to a simple model of networks with community structure. 
To this end, we actually keep the same ingredients as in our temperature-driven small-world model defined above, with the difference that, instead of assuming that the quantities $d_{ij}$ are metric distances among the nodes of a lattice, we imagine that they are \emph{ultrametric} distances among the leaves of a dendrogram.  
In simpler words, we assume that the $N$ vertices of the network can be categorized into a taxonomic tree with $N$ leaves at the bottom layer, where $d_{ij}$ is the height of the closest common branching point separating $i$ and $j$.

\subsection{Ultrametric small-world model}

Having assumed that the distances $d_{ij}$ are ultrametric, we further assume that the network is specified by eq.(\ref{eq_distance}).
The rationale for this choice is that, as an increasing body of empirical evidence suggests, the connectivity of content-rich networks such as the World Wide Web or paper citation networks is strongly determined by the semantic relationships between nodes. 
Web pages or scientific articles about similar topics are simply more likely to be connected to each other. 
Since topics are generally associated with a hierarchical taxonomy (with subcategories nested within categories), it follows that nodes with a closer common branching point in the taxonomic tree are more likely to be connected.
The above qualitative properties are precisely those featured by eq.(\ref{eq_pd}), having assumed an ultrametric distance $d_{ij}$ between all pairs of vertices. 
We can therefore regard this model as an \emph{`ultrametric small-world model'}.
A similar idea, based on the embedding of vertices in a hyperbolic space, has been found to give rise to a rich phenomenology explaining many empirical properties of real networks \cite{krioukov1,krioukov2}.

Note that, since the authors of web pages or scientific articles are typically active in specialized topics, the hierarchical taxonomy will also affect the structure of social networks defined at the level of the authors themselves, e.g. networks of authors linked by coauthorship or (co-)citation.
Moreover, even when $d_{ij}$ is defined as a `social distance' that in principle is not necessarily ultrametric (for instance, a metric distance between Euclidean vectors specifying the cultural traits of individuals), the empirical values of $d_{ij}$ are found to be approximately ultrametric \cite{mypnas}.

\begin{figure}[]
\begin{center}
\includegraphics[width=.9\textwidth]{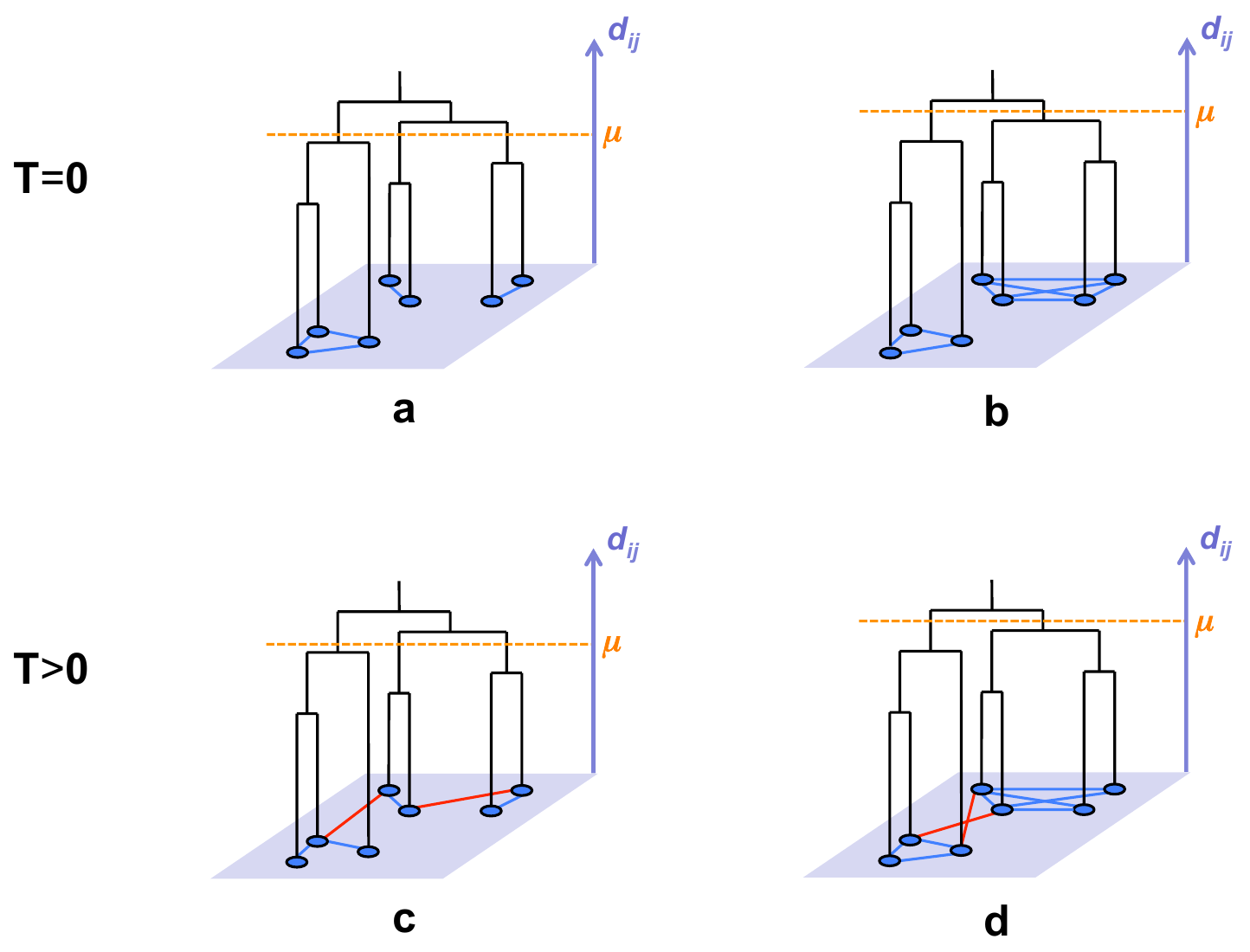}%
\end{center}
\caption{Our `ultrametric small-world model' as a function of temperature $T$ and chemical potential $\mu$. 
Nodes (blue circles) are leaves of a dendrogram (black lines), separated by an ultrametric distance $d_{ij}$ (increasing along the purple axis) representing the height of the closest branching point separating vertices $i$ and $j$.
The ultrametric distances determine the topology of the network (lying on the horizontal purple plane): \textbf{a)} when $T=0$ and $\mu$ is small, the network is divided into many small cliques (blue links) corresponding to the disconnected branches obtained by `cutting' the dendrogram along the orange dashed line determined by $\mu$; \textbf{b)} when $T=0$ and $\mu$ is larger, the network is divided into fewer and larger cliques; \textbf{c)} when $T\gtrsim 0$ and $\mu$ is small, there are many small communities that are highly connected internally (blue links) and sparsely connected across (red links); \textbf{d)} when $T\gtrsim 0$ and $\mu$ is larger, there are fewer and larger communities, with a higher density constrast between intra-community (blue) and inter-community (red) links.
After introducing an appropriate degree of heterogeneity at the level of vertices, this model can be turned into our `ultrametric scale-free model' where a non-trivial community structure coexists with a broad degree distribution.}
\label{fig:usw}
\end{figure}

In this case as well, when $T=0$ the connection probability reduces to the deterministic expression (\ref{eq:theta2}). 
However, the resulting network is now very different: the pairs of vertices separated by a distance smaller than $\mu$ are now the leaves of the dendrogram having a common branching point at a height smaller than $\mu$.
Equation (\ref{eq:theta2}) implies that all such pairs of vertices are connected, while all other pairs are not.
Visually, if we `cut' the dendrogram along the horizontal direction at a height $\mu$, all the leaves found within a connected branch will be the nodes of complete cliques (fully connected subgraphs) in the network (see fig. \ref{fig:usw}a).
Leaves belonging to different connected branches will not be connected, so the network is split into as many connected components as the number of connected branches produced by cutting the dendrogram.
Higher values of $\mu$ produce fewer and larger connected components, but at zero temperature all such components are in any case complete cliques (see fig. \ref{fig:usw}a-b).

For small but positive $T$, the zero-temperature structure will be perturbed into a finite-temperature one where the original cliques become `modules' of densely (but not completely) connected vertices, with a few links connecting different modules (see fig. \ref{fig:usw}c-d).
This is precisely the kind of community structure that is observed in most real socio-economic networks \cite{smallbutslow}.
When $T$ becomes large, more missing links will be produced within communities and more links will be produced among them, until the intra-community and inter-community densities equalize to the common value $1/2$ in the limit $T\to +\infty$. 
So, again, we find that in order to reproduce the empirical properties of socio-economic networks (where the contrast between inter- and intra-community density is very marked, but at the same time communities are not isolated from each other) the relevant regime of the model is the one found for small positive $T$.

\subsection{Ultrametric scale-free model}

However, as for the ordinary small-world model, our ultrametric variant defined above does not reproduce the broad degree distribution of real-world networks. 
But in this case as well, it is possible to introduce the desired heterogeneity at the level of vertices by extending the model as in eqs.(\ref{eq:d+e})-(\ref{eq:pDCM}), where now $d_{ij}$ is an ultrametric distance.
For a sufficiently high level of heterogeneity of $\epsilon_i$, the model preserves the scale-free character of its distance-independent counterpart described in sec. \ref{sec:fitness}. 
For this reason, we call this model the \emph{`ultrametric scale-free model'}.

It should be noted that in this case the range of variability of $d_{ij}$ can be much broader than in the non-ultrametric case, because here small intra-branch taxonomic distances coexist with large inter-branch ones.
This means that now the approximation in eq.(\ref{eq:approx}) is no longer legitimate, and we cannot reduce our model with community structure to the one without it. 
Rather, for all pairs of vertices $i$ and $j$ within the same community $C_\mu$ (as specified by $\mu$ when $T=0$) we now have the inequality
\begin{equation}
\epsilon_{ij}= d_{ij}+\epsilon_{i}+\epsilon_{j}< \mu+\epsilon_{i}+\epsilon_{j}\qquad\forall i,j\in C_\mu
\label{eq:approx2}
\end{equation}
The above expression only holds within each community, while across communities the opposite inequality applies, confirming that now distances cannot be reabsorbed in a unique value of the chemical potential $\mu$.
In other words, while our discussion in sec. \ref{sec:sw} suggested that the scale-free property automatically ensures the small-world one, here we find that the scale-free property does not automatically ensure the presence of community structure (and vice versa). Of course, when the model considered here displays a sufficiently heterogeneous degree distribution it will also automatically imply the small-world property, along the lines discussed in sec. \ref{sec:sw}.\\

We can therefore conclude that the model defined by eqs.(\ref{eq:d+e})-(\ref{eq:pDCM}), where $d_{ij}$ is an ultrametric distance, is a simple but highly nontrivial one. 
Since throughout this paper we have not been interested in reproducing a particular network, but rather a class of generic empirical properties, we will not consider any specific $d_{ij}$ but merely note that, despite its simplicity, the above model is able to reproduce all the topological properties of real socio-economic networks discussed at the beginning: the presence of a giant connected component as ensured by eq.(\ref{eq:siempre}), a strong community structure induced by the ultrametric distance $d_{ij}$, a broad degree distribution and a small-world behaviour driven by a sufficiently heterogeneous distribution of $\epsilon_i$ in eq.(\ref{eq:d+e}), and finally some realistic clustering and correlation properties as discussed in sec. \ref{sec:fitness} for the simpler model (and qualitatively preserved here).
Remarkably, for each of these properties to be there, the temperature $T$ must be small but non-zero. 
This summarizes why we believe that a small graph temperature might simultaneously explain a range of structural properties that are found ubiquitously in real social and economic networks.\\

As discussed in the Introduction, the community structure observed in real networks strongly affects the dynamics of information spreading in these systems. For most of the time, the dynamics are locally confined within communities, with occasional jumps across different communities.
At a dynamical level, this implies the so-called `small but slow world' effect, i.e. the fact that even if topologically short paths exist between most pairs of vertices, such paths are rarely explored dynamically unless they fall within one community.

The aforementioned recent study showing that the cultural distance $d_{ij}$ among individuals of a society is effectively ultrametric \cite{mypnas} has also proven that this particular distribution of individuals in cultural space has dramatic effects for both short-term and long-term information dynamics. 
In the short term, an ultrametric distance between individuals results in the onset of a `coordinated' regime where the combination of many microscopic inter-individual influences generates a society-wide collective behaviour.
In the long term however, it constrains cultural convergence to the lower branches of the dendrogram, which eventually become culturally homogeneous.
Taken together, these mechanisms were shown to explain the long-standing paradox of the coexistence of short-term collective social behaviour and long-term cultural diversity \cite{mypnas}.

These important consequences of ultrametric distances on social and cultural dynamics suggest that our `ultrametric scale-free model' promises not only to reproduce the static topology of real social networks, but to also enable a realistic simulation of distance-dependent dynamics of information diffusion.
 
\section{Weighted networks as temperature-dependent ensembles of binary graphs}

So far, we have considered binary networks. 
Even if, thanks to recent results characterizing various ensembles of weighted graphs \cite{mybosefermi,mymethod,mywreciprocity}, a full extension of our formalism to weighted networks is possible, here we only consider a simpler and practical extension that makes use of a mapping between edge weights and edge probabilities.
Quite recently, we explored the idea that the empirical weights $w_{ij}$ in a real weighted network 
can be transformed into a matrix of probabilities $p_{ij}=p(w_{ij})$ defining an ensemble of binary graphs \cite{sebsensemble}. 
Through this mapping, many topological properties which are non-obvious for weighted networks (such as the clustering coefficient) can be re-defined as ensemble averages of the corresponding binary quantities. 
In ref.\cite{sebsensemble} we explored the simplest possible choice where $p_{ij}\propto w_{ij}$. The results presented here suggest 
that this choice is the `classical' limit, equivalent to eq.(\ref{classical}), of a more general choice that we now consider.

\subsection{The temperature of real weighted networks}

Turning to eq.(\ref{eq:p}), if we require $p_{ij}=0$ when $w_{ij}=0$ and $p_{ij}=1$ when $w_{ij}=+\infty$, we find that $w_{ij}$ must be proportional to the \emph{link fitness} $e^{-\epsilon_{ij}/T}$. 
In other words, the weights must depend on $T$, which corresponds to the property that at low $T$ (more heterogeneous weights) some pairs of vertices are much more likely to be connected than other pairs, while at high $T$ (more homogeneous weights) all pairs of vertices tend to have a similar connection probability. 
Now, many real networks \cite{flux,interbank,another,kim,burda_fin} display a power-law distribution of \emph{non-zero} link weights of the form 
\begin{equation}
\rho(w)\propto w^{-\alpha}\qquad 1.5\lesssim\alpha\lesssim 3.5
\end{equation}
If we restrict ourselves to pairs of vertices with $w_{ij}>0$ and define $x_{ij}\equiv w_{ij}/w_{min}\ge 1$ (where $w_{min}$ is the minimum non-zero weight for a given network), corresponding to the preliminary rescaling $\epsilon_{ij}\to\epsilon_{ij}-\epsilon_{max}$, then we can repeat the arguments leading to eq.(\ref{eq_Tdistr}). Specifically, we set $\epsilon_{ij}\equiv-T\ln x_{ij}\le 0$ and $\phi_{ij}\equiv-\epsilon_{ij}\ge 0$ to obtain
\begin{eqnarray}
q(\phi)=e^{-\phi}\quad (\phi\ge 0)\label{eq_phidistr2}\\
\rho(x)=Tx^{-1-T}\quad(x\ge 1)
\label{eq_Tdistr2}
\end{eqnarray}
where now $\rho(x)$ and $q(\phi)$ are distributions not over vertices, but over pairs of them (specifically, over the pairs with non-zero weights).
This allows us to compute the temperature of real networks with power-law distributed weights  as 
\begin{equation}
T_{weighted}=\alpha-1\quad\implies \qquad 0.5\lesssim T_{weighted}\lesssim 2.5
\end{equation}
The empirical values of $\alpha$ found in various weighted networks \cite{flux,interbank,another,kim,burda_fin} are summarized in Table \ref{tab}, and the corresponding values of $T_{weighted}$ are also shown. By contrast, note that binary networks (where all weights are equal) correspond to  $T\to\infty$ where $x_{ij}=1$ $\forall i,j$.\\

We have therefore found that a general mapping from weights to probabilities is given by \begin{equation}
p_{ij}=\frac{zx_{ij}}{1+zx_{ij}}
\label{eq:EN}
\end{equation}
where $x_{ij}\equiv w_{ij}/w_{min}$ and $z\equiv e^{\mu/T}$ is a free parameter. 
Note that the above expression works for both zero and non-zero weights.
We also note that the classical limit (\ref{classical}) of this expression reads $p_{ij}\approx zx_{ij}$, and if we choose $z=w_{min}/w_{max}$ we have \begin{equation}
p_{ij}\approx \frac{w_{ij}}{w_{max}}
\end{equation}
which is approximately equivalent to the choice explored by us in ref.\cite{sebsensemble}. 

\begin{table}[t]
\begin{centering}
\begin{tabular}{|l|c|c|c|}
\hline
Network & $\alpha$ & $T_{weighted}$ & Ref. \\
\hline
Metabolic flux networks & $1.5$ & $0.5$ & \cite{flux}\\
Interbank network & $1.87$ & $0.87$&\cite{interbank}\\
Erd\H{o}s collaboration network & $2$ & $1$ &\cite{another}\\
Chaos control \& synchron. coauthorship & $2.5$ & $1.5$ & \cite{another}\\
Financial cross-correlations & $2.7$ & $1.7$ & \cite{kim}\\
Financial cross-correlations & $2.78$ & $1.78$ & \cite{burda_fin}\\
Financial cross-correlations & $3.18$ & $2.18$ & \cite{burda_fin}\\
Mollusk research coauthorship & $3.5$ & $2.5$ & \cite{another}\\
Binary graphs & $+\infty$ & $+\infty$ & \\
\hline
\end{tabular}
\caption{Empirical values of $\alpha$ and $T_{weighted}$ for some real weighted networks.}
\label{tab}
\end{centering}
\end{table}

\subsection{Filtering of weighted networks as the zero-temperature limit}

When $T\to 0$, the expression (\ref{eq:EN}) defining our generalized mapping from a weighted network to an ensemble of binary graphs reduces to eq.(\ref{eq:theta}). This implies that the original weighted network is mapped into a deterministic binary one where only the links with $\epsilon_{ij}<\mu$ are drawn. This means that the links 
with weight such that $x_{ij}(T)>z^{-1}(T)$ in the limit $T\to 0$ are selected and the others are discarded. 
Interestingly, since the ordering of the weights is preserved at all temperatures, this corresponds to a standard thresholding procedure, adopted for instance in ref.\cite{stocks} to filter stock correlations and in ref.\cite{foodwebs} to extract minimum spanning trees from real foodwebs. 
These filtering techniques discard most of the information contained  
in the weights, resulting in a single (threshold-dependent)  
binary graph. Here we find that this corresponds to the zero-temperature limit of our mapping from weighted networks to ensembles of binary graphs. 
Our results extend these techniques to the finite temperature case, making it possible to preserve the heterogeneity of the links and explore the whole ensemble of possible configurations with the appropriate probabilities.
We expect that this will represent an improved filtering technique, with a significantly reduced information loss.

\section{Conclusions}

We have introduced the concept of `graph temperature', which can vary from zero to infinity, in order to explore the behaviour of networks in the limit of large network size while keeping the local properties well-defined. 
Since our methodology makes use of statistical graph ensembles that extend the class of Exponential Random Graphs widely used in social network analysis, it has a natural application as a generalized model of social and economic networks.
We showed that many structural properties that are ubiquitous in socio-economic networks can be simply understood as the effects of an optimized low-temperature behaviour resulting from `connectivity costs', and confirmed this by measuring the temperature of both binary and weighted real-world scale-free networks.
Furthermore we have also shown that a variety of different models and techniques can in fact be regarded as particular cases of a more general temperature-dependent formalism.
We believe that our results provide an intuitive and unified understanding of many properties of real socio-economic networks, from their scale-free and small-world behaviour to their hierarchical community structure.

\acknowledgements{Acknowledgements}

D.G. acknowledges support from the Dutch Econophysics Foundation (Stichting Econophysics, Leiden, the Netherlands) with funds from beneficiaries of Duyfken Trading Knowledge BV, Amsterdam, the Netherlands.
S.E.A. acknowledges support from The Leverhulme Trust, UK and The Royal Society, UK.
G.C. acknowledges support from FET project FOC (255987) and MULTIPLEX (317532).


\conflictofinterests{The authors declare no conflict of interest.}

\bibliographystyle{mdpi}
\makeatletter
\renewcommand\@biblabel[1]{#1. }
\makeatother


\end{document}